\documentclass[twocolumn,a4paper,footinbib,
showpacs,aps,floatfix]{revtex4}

\usepackage{graphicx}
\usepackage{bm}
\usepackage{amsmath}
\def\la{\langle}
\def\ra{\rangle}

\begin {document}
\title{Vibronic ``Rabi resonances'' in harmonic and hard-wall 
ion-traps for arbitrary laser intensity and detuning}
\author{I. Lizuain}
\affiliation{Departamento de Qu\'\i mica-F\'\i sica,
Universidad del Pa\'\i s Vasco, Apdo. 644, Bilbao, Spain}
\author{J. G. Muga}
\affiliation{Departamento de Qu\'\i mica-F\'\i sica,
Universidad del Pa\'\i s Vasco, Apdo. 644, Bilbao, Spain} 


\begin{abstract}
We investigate laser-driven vibronic transitions of a single two-level 
atomic ion in harmonic and hard wall traps.
In the Lamb-Dicke regime, for tuned or detuned lasers with respect to the 
internal frequency of the ion, and weak or strong laser intensities,  
the vibronic transitions 
occur at well isolated ``Rabi Resonances'', 
where the detuning-adapted Rabi frequency 
coincides with the transition frequency between 
vibrational modes. These vibronic resonances are characterized as avoided crossings 
of the dressed levels (eigenvalues of the full Hamiltonian). 
Their peculiarities due to symmetry constraints and trapping potential are 
also examined. 
\end{abstract}
\maketitle
%
%
%
%
\section{Introduction}
Laser cooled trapped ions have been much studied, both theoretically and 
experimentally, because of metrological applications as frequency standards,  
high precision spectroscopy, or the prospects of realizing quantum information processing \cite{review}. 
The coupling between internal and vibrational degrees of freedom
of the ion in harmonic traps  
due to laser light has in particular been examined as a way to implement the Jaynes-Cummings model or other simple model Hamiltonians for quantum gate 
design, motional cooling or state preparation \cite{CBZ94,CZ95,VM95}. This coupling
has been predominantly considered for  
weak fields, $\Omega_R<<\omega_T$, where $\omega_T$ is the trap's lowest (angular) frequency (for direction $x$, we assume that the two other frequencies in $y$ and $z$ directions are so large that the ion motion is effectively one dimensional),  and $\Omega_R$ is the optically ``on-resonance''  
Rabi frequency, i.e., for the zero-detuning case in which the laser frequency $\omega_L$ and the ion transition  frequency $\omega_0$ between two internal levels coincide, 
$\Delta:= \omega_L-\omega_0=0$ 
\footnote{ 
The laser frequency, and laser wavenumber are effective ones if a two photon Raman transition
is used. $\Omega_R$ is assumed to be real.}.
In the laser adapted interaction picture, after applying the usual (optical) rotating wave approximation and dropping the constant term of the zero point energy,
the Hamiltonian reads for a classical laser field,  
\begin{equation}
\label{H}
H=H_{trap}+H_A+V(\Omega_R),
\end{equation}
where 
\begin{eqnarray}
H_{trap}&=&\hbar\omega_T a^{\dag}a,
\\
H_A&=&-\frac{\hbar\Delta}{2}\sigma_z,
\nonumber\\
V&=&\frac{\hbar\Omega_R}{2}\left[
e^{i\eta\left(a+a^\dag\right)}
|e\ra\la g|
+ h.c\right],
\end{eqnarray}
$\eta:=k_L x_0$ is the Lamb-Dicke (LD) parameter,
with $x_0$ being the characteristic length of the oscillator
$x_0=\sqrt{\frac{\hbar}{2m\omega_T}}$, and $k_L$ the 
laser wavenumber; also        
$\sigma_z=|e\ra\la e|-|g\ra\la g|$, 
and $a$ and $a^\dag$ are the annihilation and 
creation operators for the vibrational states $\{|n\ra\}, n=0,1,2,...$. 
We are assuming that the internal states $\{|g\ra, |e\ra\}$
of the ion have an infinite lifetime, thus neglecting spontaneous emission.

At this point it is customary to 
separate the Hamiltonian as \cite{Orszag}
\begin{eqnarray}
\label{Homega}
H&=&H_B+V(\Omega_R),
\\
H_B&=&H_{trap}+H_A,
\end{eqnarray}
to carry out a second interaction picture
with respect to $H_B=H(\Omega_R=0)$, i.e., the free motion of the electronic and 
vibrational degrees of freedom of the ion or ``bare part''.     
If the rapidly oscillating  
terms are neglected in a ``second'' or vibrational rotating wave approximation
and, in addition,  
the Lamb-Dicke regime is assumed, $\eta<<1$, 
three possible resonance   
conditions arise between the detuning and the vibrational frequency. 
They define motional sidebands and state subspaces with simplified, 
effective Hamiltonians:     
\begin{itemize} 
\item $\Delta=0$:  ``carrier'' sideband,   
a purely internal resonance between bare states
$|g,n\ra$ and $|e,n\ra$, without vibrational excitation. 
\item $\Delta=-\omega_T$: first red sideband for vibronic transitions between $|g,n+1\ra$ and $|e,n\ra$ represented by a Jaynes-Cummings model Hamiltonian. 
\item $\Delta=\omega_T$: first blue sideband for 
vibronic transitions 
between $|g,n\ra$ and $|e,n+1\ra$ (anti-Jaynes-Cummings model).
\end{itemize}
These transitions are the basis of the Cirac-Zoller and related controlled-NOT quantum gates \cite{CZ95,Wineland95,Nature}.     
By choosing larger detunings, one can obtain in principle further  
vibronic resonances (motional sidebands) with non-linear couplings 
depending on higher orders of the Lamb-Dicke parameter, but they are
not easy to realize in the laboratory since efficient motional cooling is usually
performed in the Lamb-Dicke regime \cite{review}. 

The vibrational rotating wave approximation breaks down for larger 
$\Omega_R$. This is but a particular example of the failure of the ``secular 
approximations'' for a strong perturbation \cite{Cohen}.
There are however good reasons for exploring also stronger fields since     
one of the consequences of weak fields is that the vibronic transitions involving 
one vibrational quantum occur slowly: the effective Rabi frequencies for them 
scale with $\eta\Omega_R$. Looking for  faster quantum gates, Jonathan, Plenio and Knight \cite{JPK00} considered the possibility of using instead stronger fields for the optically resonant case, $\Delta=0$, finding a different vibronic resonance condition, $\Omega_R=\omega_T$, for transitions involving one motional quantum and 
laser adapted internal states $|\pm\ra=(1/\sqrt{2})(|g\ra\pm|e\ra)$.   
These transitions had been already noticed by Moya-Cessa et al. \cite{MC99}  in a study of super-revivals, by means of a unitary transformation 
leading to the transformed Hamiltonian  
\begin{eqnarray}
\label{HMC}
H_{MC}&=&H_{trap}
+\frac{\hbar\Omega_R}{2}\sigma_z
\nonumber\\
&+&\!\hbar\frac{\Delta}{2}(\sigma_-\!+\!\sigma_+)
\!+\!\frac{i\hbar\eta\omega_T}{2}(a\!-\!a^\dagger)(\sigma_-\!+\!\sigma_+), 
\label{hmc}
\end{eqnarray}
where $\sigma_-=|g\ra\la e|$ and $\sigma_+=|e\ra\la g|$. 
Going to an interaction picture with respect to $H_{trap}+\frac{\hbar\Omega_R}{2}\sigma_z$, and applying again 
a rotating wave approximation to get rid of rapidly oscillating terms gives the 
vibronic  
resonance condition $|\Omega_R|=\omega_T$ and suppresses any dependence on the detuning, but, mirroring  the previously 
commented  failure of the RWA 
for ``large'' $\Omega_R$, this scheme is clearly not valid for sufficiently 
large detuning $\Delta$, and requires $\Delta<<\omega_T$.
As pointed out in \cite{APS03}, in the weak field limit the  
last term (``spin flip term'') in Eq. (\ref{hmc}) does not go to zero so it does not provide a good perturbative scheme. Note also that higher, nonlinear resonances, of the type 
$\Omega_R=k\omega_T$, $k=\pm 2, \pm 3, ...$, are not immediately 
obvious using a second rotating wave approximation although, in principle, could be identified with a 
more sophisticated treatment \cite{KSC03}.   

Aniello, Porzio, and Solimeno \cite{APS03} could go beyond these
two partial approaches, valid for weak fields or weak detunings, and set a 
more general  resonance condition for the principal vibronic resonance,
namely for $\Omega=\omega_T$, where 
\begin{equation}
\Omega=(\Omega_R^2+\Delta^2)^{1/2}\ge 0
\end{equation}
is the detuning-adapted Rabi frequency \footnote{Note that our notation convention and terminology may differ from the quoted papers. For example, 
our $\Omega_R$ is twice the corresponding quantity in \cite{JPK00}, \cite{MC99}, or \cite{APS03}, the detuning is defined quite often with opposite sign, or  
our Rabi frequency $\Omega$ is called ``corrected detuning'' in \cite{APS03}}. 
Their treatment, however, required a somewhat complicated series of transformations
and relied on finding a zeroth order Hamiltonian diagonal in the bare basis 
of the bare Hamiltonian $H_B=H(\Omega_R=0)$. 
We shall present here a simpler treatment  in which 
the initial Hamiltonian (\ref{H}) is separated into 
the following zeroth order and perturbation terms 
\begin{equation}
\label{ourH}
H=H_{SD}+W(\eta),
\end{equation}
where the ``semidressed'' Hamiltonian is given by $H(\eta=0)$, 
\begin{equation}
H_{SD}=\hbar\omega_T a^\dag a -\frac{\hbar\Delta}{2}\sigma_z +
\frac{\hbar\Omega_R}{2}\left(|e\ra\la g| +|g\ra\la e|\right),
\end{equation}
and represents the ion in the trap with a laser field which does not induce any 
motional coupling,
whereas  
\begin{equation}
\label{w_perturbation}
W(\eta)=\frac{\hbar\Omega_R}{2}\left[ \left( e^{i\eta(a+a^\dag)}-\boldsymbol {1}\right)|e\ra\la g|+h.c\right]
\end{equation}
represents the coupling term. The explicit dependence in the argument of $W(\eta)$ emphasizes that
$H=H_{SD}$ when $\eta=0$.
Notice that the leading term in Eq. (\ref{w_perturbation}) 
depends on $\eta\Omega_R$, whereas 
$H_{SD}$ depends parametrically on $\omega_T, \Delta$ and $\Omega_R$. This 
will determine the parameter domains appropriate for a perturbative scheme based on Eq. (\ref{ourH}). 
All vibronic resonances discussed so far for weak field, or weak detuning  
will be identified directly from the degeneracies of the ``semidressed'' eigenenergies of $H_{SD}$.
With the coupling term $W(\eta)$, they become avoided crossings and their   
width and energy splitting are well characterized  
by perturbation theory.       
In this way, the vibronic resonance 
condition of Aniello et al. may be obtained and generalized from the crossings of the energy levels of a zeroth order Hamiltonian in the whole range of $\Omega_R$ and $\Delta$ values. Since the vibronic resonances correspond to Rabi frequencies equal to the vibrational transition frequencies, the term ``Rabi resonance'', 
which we proposed previously for a related waveguide confinement \cite{LRM06},
seems appropriate for the general case. 

For small Lamb-Dicke parameters (the smallness criterion depends on the particular resonance as we shall see), the resonances are well isolated, which facilitates 
state manipulation and control.  
Other factors affecting any possible use of the resonances for information processing 
or state preparation are the speed of the transition, determined by the effective
Rabi frequency at the vibronic resonance, and its width, which will tell us the minimum stability 
required in the laser detuning and/or intensity. 
We shall see that both factors as well as the isolation  
of the resonances can be deduced from the 
energy splitting, for which we shall provide explicit and resonance-dependent expressions.  

The formalism is also easy to apply for a hard-wall trap, and comparisons are 
made with the harmonic one.
The hard-wall trap has been experimentally realized optically \cite{MSHCR05} or  in a microelectronic chip
version \cite{HHHR01}, and has generated recently much theoretical work in
the field of ultracold atoms in low dimensions
\cite{Gaudin71,Cazalilla02, BGOL05,DM05b,mbd}.

In the next section we shall first rephrase the weak field case 
in a time independent framework following \cite{CBZ94}. This will be 
a useful reference since the more general case of arbitrary laser intensity in Sec.  III 
can be treated very similarly. In Section IV we compare the results 
for harmonic and square traps and the paper ends with a discussion and technical appendices.

%
%
%
%
%
%
%
%

\section{Low intensity fields; time independent approach}
\label{low_intensity}
%
\begin{figure}[t]
\includegraphics[height=6cm]{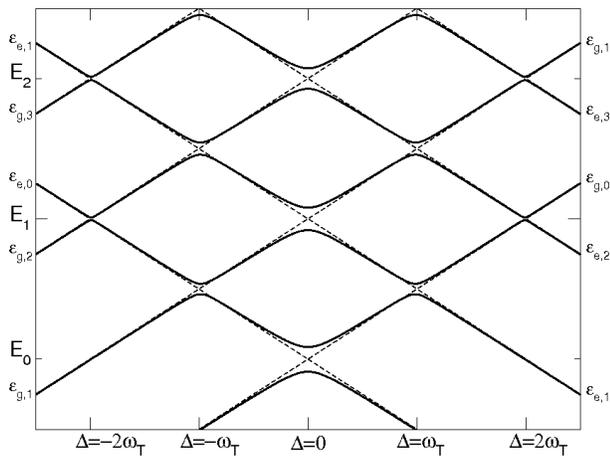}
\caption[]{Bare ($\Omega=0$, dashed line) and dressed ($\Omega/\omega_T=0.2$, solid line) energy levels in the resolved
sideband regime as a function of the laser detuning. The 
Lamb-Dicke parameter is $\eta=0.4$ in both cases. Note the degeneracies in the bare case at $\Delta=\pm k\omega_T$ become
avoided crossings when the states are dressed ($k^{th}$ motional sidebands).  }
\label{bare_levels}
\end{figure}
%
In this section we present a time independent approach \cite{CBZ94} which is essentially equivalent to the usual time-dependent treatment based on a second application of the RWA  \cite{Orszag,review}.
For $\Omega_R\ll\omega_T$ \cite{CBZ94,Vogel},
it is natural to regard the coupling between the laser and the ion as 
a small perturbation of the bare system consisting on the trapped ion, and 
decompose the Hamiltonian
(\ref{H}) as in Eq. (\ref{Homega}),  
$H=H_{B}+V(\Omega_R)$. 
%
%
%
The bare part, $H_B=H(\Omega_R=0)$,  is diagonal in the bare basis $\lbrace |g,n\ra,|e,n\ra\rbrace$ with eigenvalues 
\begin{eqnarray}
\epsilon_{g,n}&=&E_n+\frac{\hbar\Delta}{2},\\
\epsilon_{e,n}&=&E_n-\frac{\hbar\Delta}{2},
\end{eqnarray}
where $E_n=n\hbar\omega_T$ are the energy levels of the oscillator.
These bare energy levels cross each other whenever the detuning is on-resonance with the vibrational-level spacing,
see Fig. \ref{bare_levels} (dashed line), 
\begin{equation}
\label{delta_res_con}
\Delta=k\omega_T,
\end{equation}
with $k=0,\pm 1,\pm 2,\hdots$.
When the laser is applied, these degeneracies are removed and the crossings become avoided crossings, see
Fig. \ref{bare_levels} (solid line).
This is just a particular case of the ``no level crossing theorem'', which
states that a pair of energy levels connected by a perturbation do not 
cross as the strength of the perturbation is varied. The matrix elements
of the coupling potential are 
\begin{eqnarray}
\la g,n|V(\Omega_R)|e,n'\ra&=&\frac{1}{2}\hbar\Omega_{nn'}^{*},\\
\la e,n|V(\Omega_R)|g,n'\ra&=&\frac{1}{2}\hbar\Omega_{nn'},
\end{eqnarray}
(the rest of terms are zero), 
where the effective Rabi frequency $\Omega_{nn'}$, defined by
\begin{equation}
\label{rabi_frequencies}
\Omega_{nn'}:=\Omega_R \la n|e^{i\eta(a+a^\dag)}|n'\ra=\Omega_{n'n}
\end{equation}
is the coupling strength between the motional levels $n$ and $n'$,  
calculated in Appendix \ref{rabi_calculation}.
These matrix elements are in general different from zero, so that
all the ground and excited levels are connected by the perturbation and therefore do not cross.
Moreover, these avoided crossings will be well-localized and isolated, because of  the assumption of a low intensity laser.
Good approximations to the dressed states at avoided crossings will be obtained by diagonalizing the $2\times2$ reduced matrix in the degenerate subspace.
If we consider the crossing of the $|g,n\ra$ and $|e,n'\ra$ levels at the $\Delta=\frac{E_{n'}-E_n}{\hbar}=k\omega_T$ ($k=n'-n$) resonance, 
the matrix in the degenerate subspace spanned by $|g,n\ra$ and $|e,n'\ra$ reads
\begin{equation}
\label{2matrix_resolved_sideband}
\tilde V_{\lbrace|g,n\ra,|e,n'\ra\rbrace}=\frac{\hbar}{2}\left(\begin{array}{cc}
0&\Omega_{n'n}^{*}\\
\Omega_{nn'}&0
\end{array}\right).
\end{equation}
This operator governs the dynamics of the system at the resonance, 
leading to Rabi Oscillations 
with a frequency ${|\Omega_{nn'}|}$ between bare states $|g,n\ra\leftrightarrow|e,n'\ra$.
These resonances are nothing but the $k^{th}$  motional 
sidebands mentioned before \cite{review}. 

The eigenvalues of the matrix (\ref{2matrix_resolved_sideband}) provide the energy splitting in the avoided
crossing, 
\begin{equation}
\label{split}
\Delta\epsilon_{nn'}=\hbar|\Omega_{nn'}|.
\end{equation}
In leading order (see Appendix \ref{rabi_calculation}), 
$\Delta\epsilon_{nn'}\propto \Omega_R \eta^{|n-n'|}$. 
A criterion for resonance isolation is $\Delta\epsilon_{nn'}<<\hbar\omega_T$, since  
$\hbar\omega_T$ is the energy difference between consecutive resonances.
This leads to the condition $\Omega_R<<\omega_T$ for the carrier, 
and $\eta\Omega_R<<\omega_T$ for the first sidebands, which leaves some room 
for a relatively high $\eta$ beyond the LD regime \cite{CBZ94}.    
%
%
%
%
%
\section{Arbitrary intensity fields}
If more intense lasers are to be used, e.g. for $\Omega_R\gtrsim\omega_T$, 
the resonances described before 
are not isolated and new transitions with different sidebands occur. 
It is possible however, to find well isolated avoided crossings
for higher intensity lasers if we work in the Lamb-Dicke regime 
where it is more 
natural to regard the LD parameter $\eta$ as the perturbative parameter, 
as in Eq. (\ref{ourH}), $H=H_{SD}+W(\eta)$, instead of the on-resonance Rabi frequency used in 
the decomposition (\ref{Homega}). 
%
%
%

The semi dressed Hamiltonian $H_{SD}$ is easily diagonalized, with eigenvalues and corresponding energy eigenstates
\begin{eqnarray}
\label{semi_dressed_levels}
\epsilon_{n,\pm}&=&E_n\pm\frac{\hbar\Omega}{2},
\label{slopes}\\
|\epsilon_{n,\pm}\ra&=&\frac{1}{\sqrt{N_{\pm}}} \left( \frac{\Delta\pm\Omega}{\Omega_R}|g,n\ra+|e,n\ra\right),
\label{semi_dressed_states}
\end{eqnarray}
where the $N_{\pm}$ are dimensionless
normalization factors, 
\begin{equation}
\label{normalization}
N_{\pm}=\frac{\left(\Delta\pm\Omega\right)^2}{\Omega_R^2}+1=\frac{2\Omega}{\Omega_R^2}\left(\Omega\pm\Delta\right). 
\end{equation}
These semidressed energy levels are degenerate at ``Rabi Resonances", where the detuning-adapted Rabi frequency is resonant with
some vibrational transition,
\begin{equation}
\label{rabi_resonance_condition}
\Omega=\frac{|E_{n'}-E_n|}{\hbar}=|n'-n|\omega_T,
\end{equation}
or $\Omega_R^2+\Delta^2=(n'-n)^2\omega_T^2$. 
This also provides the motional sideband resonances in the
low intensity limit ($\Omega_R\ll\omega_T,\Delta$): the semidressed states in Eq. (\ref{semi_dressed_states}) reduce to the bare states and are 
resonant at $\Delta=(n'-n)\omega_T$.
A peculiarity of the present approach compared to the one in Sec. II is that, since $\Omega$ is defined as a positive number,
red and blue sidebands are formally treated in the same manner, 
see also \cite{APS03} \footnote{The distinction  occurs because the 
states $|\epsilon_{n,\pm}\ra$ in Eq. (\ref{semi_dressed_states}) tend to 
$|g,n\ra$ or $|e,n\ra$ depending on the sign of $\Delta$.}.  
%
%

\begin{figure}[t]
\includegraphics[height=6cm]{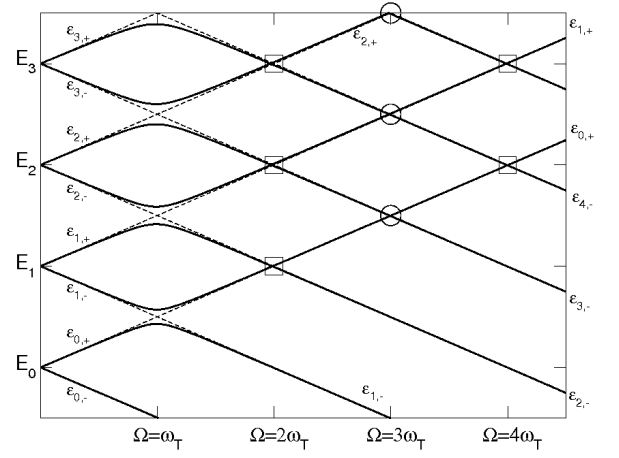}
\caption[]{Semidressed ($\eta=0$, dashed line) 
and dressed ($\eta=0.1$, solid line) energy levels for the resonant $\Delta=0$ case as a function of the 
Rabi frequency $\Omega$. 
Note the degeneracies in the semidressed case at $\Omega=|n'-n| \omega_T$ are only removed when dressed
for odd values of $k=n'-n$. Degeneracies surrounded by squares are real crossings of the energy levels, while the
circles represent avoided crossings (which are not evident in the scale of the figure).}
\label{semi_dressed_levels_fig}
\end{figure}

At Rabi Resonances, the semidressed states involved in the crossing are coupled. 
The strength of this coupling is given by
\begin{eqnarray}
\label{matrix_elements}
&&\la\epsilon_{n,s}|W(\eta)|\epsilon_{n',s'}\ra=\nonumber\\
&=&
\frac{\hbar}{2}\frac{1}{\sqrt{N_sN_{s'}}}\left[(\Delta+s'\Omega)\la n|e^{i\eta(a+a^\dag)}-1|n'\ra\right.\nonumber\\
&+&\left.(\Delta+s\Omega)\la n|e^{-i\eta(a+a^\dag)}-1|n'\ra\right]
\\
&=&\frac{\hbar\Delta}{\sqrt{N_sN_{s'}}}\left[C_{nn'}-\delta_{nn'}\right]\nonumber\\
&+&s'\frac{\hbar\Omega}{\sqrt{N_sN_{s'}}}
\left\{ \begin{array}{cc}
C_{nn'}-\delta_{nn'} & \textrm{if $s=s'$}\nonumber\\
\hspace{1cm}\\
iS_{nn'} & \textrm{if $s\ne s'$}
                         \end{array}\right.
\end{eqnarray}
where $s=\pm 1$ is a shorthand notation to represent the ``sign'' of each state and 
\begin{eqnarray}
C_{nn'}&=&\la n|\cos\eta(a+a^\dag)|n'\ra,\\
S_{nn'}&=&\la n|\sin\eta(a+a^\dag)|n'\ra.
\end{eqnarray}
%
From parity arguments it is clear that, generically, 
all these coupling elements may be different from zero, connecting all the semidressed states among them.
For the $\Delta=0$ case, however, only states with different vibrational parity will be connected, see 
Fig. \ref{semi_dressed_levels_fig}, where only alternate crossings are avoided.

The eigenvalues of the $2\times2$ reduced matrix in the degenerate subspace, 
\begin{equation}
\label{2times2_gral}
\hat W_{n,n'}=
\left(\begin{array}{cc}
\la \epsilon_{n,+}|W|\epsilon_{n,+}\ra&
\la \epsilon_{n,+}|W|\epsilon_{n',-}\ra
\\
\la \epsilon_{n',-}|W|\epsilon_{n,+}\ra&
\la \epsilon_{n',-}|W|\epsilon_{n',-}\ra
\end{array}\right),
\end{equation}
%
provide the energy splitting in the  
avoided crossing,
\begin{eqnarray}
&&\Delta\epsilon_{nn'}=\frac{\hbar\Omega_R}{2\Omega}
\\
&\times&\!\sqrt{\Omega_R^2\left(C_{n'n'}\!-\!C_{nn}\right)^2\!+\!4\left(\Delta^2 C^2_{nn'}\!
+\!\Omega^2 S^2_{nn'}\right)}\nonumber,
\end{eqnarray}
%
%
%
with leading order in $\eta$ (see Appendix \ref{rabi_calculation})
\begin{eqnarray}
\Delta\epsilon_{n,n+l}=
\hbar\Omega_R\eta^l\frac{\sqrt{(n+l)!/n!}}{l!}\left\{ \begin{array}{cc}
\frac{\Delta}{\Omega}& \textrm{if $l$ even}\\
\hspace{.1cm}\\
1& \textrm{if $l$ odd}
\end{array}\right..
\label{harmo}
\end{eqnarray}
%

%
%
In particular, 
%
%
%
%
%
%
$\Delta\epsilon_{n,n+1}=\hbar\eta\Omega_R\,\sqrt{n+1}$. 
%
Keep in mind that $\Omega_R$ in these expressions has to satisfy the resonance condition
(\ref{rabi_resonance_condition}).  

The criterion for isolated resonances is $\Delta\epsilon_{n,n+l}<<\hbar\omega_T$
whereas the resonance width with respect to $\Omega$ 
can be estimated from the slopes in Eq. (\ref{slopes}) to be  
given also by the energy splitting. 

All vibronic transitions in Sec. II can be
described in this manner, in particular we recover  
the splitting of Eq. (\ref{split}) for $\Delta>>\Omega_R$ in leading order. At the sideband-motion resonances 
for $k=\pm 1, \pm 2,...$, $\Delta$ is a multiple of $\omega_T$ 
so this condition is well satisfied.         
The energy splitting of the carrier sideband,
a purely internal transition, is also recovered in that limit, 
but it is clear that this resonance does not require $\Delta>>\Omega_R$. 
In fact in Section II, the expression for the carrier splitting is obtained for $\Delta=0$ and a small, finite 
$\Omega_R<<\omega_T$. Notice that in this regime $H(\eta=0)$ already provides 
the carrier transition splitting, so this regime lies outside the range 
of our perturbative scheme based on $W(\eta)$ breaking the degeneracy.      

For the first vibronic resonance $n=0, n'=1$ at $\Delta=0$ the resonance condition 
is $|\Omega_R|=\omega_T$ which implies a splitting
$\hbar \eta \omega_T$ 
and an isolation condition $\eta<<1$ equivalent to 
the LD condition. The isolation condition is therefore more stringent on $\eta$ for these resonances than for the motional weak-field sidebands described in Sec. II.     
    
Another difference is the dependence of the effective Rabi frequency 
and the isolation condition on $\omega_T$. Let's recall that, for a given laser 
wavenumber, $\eta\propto \omega_T^{-1/2}$. Thus $\Omega_R\eta<<\omega_T$ 
is satisfied better and better for increasing $\omega_T$, i.e., for tighter 
confinement irrespective of the type of resonance: for small $\Omega_R$
as in Sec. II, or for 
small $\Delta$. However, the splitting, and thus the speed of the transition, 
decreases with increasing $\omega_T$ in the first case, whereas 
it increases in the second, because $\Omega_R\approx \omega_T$.

\begin{figure}[t]
\includegraphics[height=6cm]{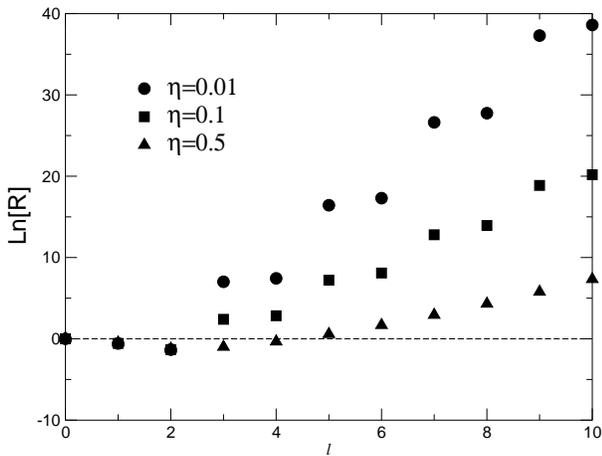}
\caption[]{Logarithm of the ratio $R:=|\chi_{_{1,1+l}}/\Omega_{_{0,0+l}}|$ 
for the first few sidebands (starting from the ground state) and for different
values of the LD parameter $\eta$. Note that for the first and second sidebands
the coupling strength is larger in the harmonic potential, 
while for higher order multi-phonon transitions, coupling is stronger in the hard-wall trap.}
\label{ratios}
\end{figure}
%
%
%
%
%
%
%
%
%
%
%
\section{Hard-wall trap}

We now consider the confinement of the ion in a hard-wall trap
\begin{eqnarray}
V(x)=
\left\{ \begin{array}{cc}
0 & \textrm{if $|x|<\frac{a}{2}$}\nonumber\\
\hspace{1cm}\\
\infty & \textrm{if $|x|>\frac{a}{2}$}
\end{array}\right..
\end{eqnarray}
The treatment is essentially the same as for the harmonic trap, 
replacing $\exp[i\eta(a+a^\dag)]$ by $\exp(ik_Lx)$, and  
the energies and 
eigenstates of the harmonic oscillator by those of the square 
potential. Now the motional energies   
\begin{equation}
E_n=\frac{\hbar^2}{2m}\left(\frac{n\pi}{a}\right)^2,
\end{equation}
%
with $n=1,2,\hdots$ are no longer equally spaced. Note that the ground state in this trap is labelled with $n=1$ in 
contrast to the ground $n=0$ state of the harmonic potential.
The hard-wall coupling strengths between the different motional levels 
(playing the role of $\Omega_{nn'}$ in the harmonic wall) are  
\begin{equation}
\chi_{nn'}=\Omega_R\la \phi_n|e^{ik_L x}|\phi_{n'}\ra=\Omega_R\!\int_{-\frac{a}{2}}^{\frac{a}{2}}\!dx \phi_n(x) \phi_{n'}(x) e^{ ik_Lx},
\end{equation}
which are explicitly calculated in Appendix \ref{square_potential}. 

%
The crossing of the semidressed energy levels define the resonances as in the harmonically trapped ion.
In the case of a hard-wall trap,
the Rabi Resonance condition is given by \cite{LRM06}
\begin{equation}
\Omega=\frac{|E_n-E_{n'}|}{\hbar}=\frac{\hbar\pi^2}{2ma^2}|n^2-{n'}^2|.
\end{equation}
%
%

\begin{figure}[t]
\includegraphics[height=6cm]{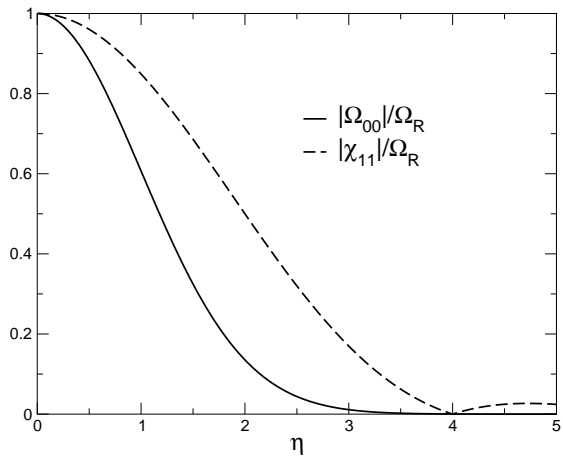}
\caption[]{Relative coupling strenght for a carrier $|g,n_g\ra\leftrightarrow|e,n_g\ra$ transition
in a harmonic ($n_g=0$, solid line) and in a hard-wall trap ($n_g=1$, dashed line) as a function of the LD parameter $\eta$.}
\label{carrier}
\end{figure}

%
The energy splitting in a given Rabi Resonance
will be given by the same formal expressions as for the 
harmonic case, replacing $\Omega_{nn'}$ by $\chi_{nn'}$.
In leading order in $\eta$, defined now as $k_L a/\pi$, 
\begin{eqnarray}
\Delta\epsilon_{n,n+l}=\hbar \frac{4\Omega_Rn(n+l)}{l^2(2n+l)^2}
\left\{ \begin{array}{cc}
\eta^2\frac{\Delta}{\Omega}& \textrm{if $l$ even}\\
\hspace{1cm}\\
\frac{2\eta}{\pi}& \textrm{if $l$ odd}
                        \end{array}\right.,
\end{eqnarray}
which should be compared with Eq. (\ref{harmo}). 
Different dependencies on the LD parameter $\eta$ are observed. 
Note that for zeroth, first and second sidebands ($l=0,1,2$), both splittings are of the same order, while for higher 
order multi-phonon transitions ($l>2$), transitions in the rigid wall potential become faster if a strong confinement
is assumed (LD regime), see Fig. \ref{ratios}.
In fact numerical comparison shows that the carrier transition is also stronger
in a hard wall for typical values of the LD parameter, see Fig. \ref{carrier}.

\section{Discussion}
The perturbative schemes based on 
different decompositions of the Hamiltonian 
$H$ in Eq. (\ref{H}), to treat the effects of a 
laser on an ion in a harmonic trap, lead naturally to the identification of 
different families of vibronic resonances, i.e., transitions 
localized in a relatively small range of some varying parameter 
(such as the detuning, the laser intensity, a combination of both, 
or even the Lamb-Dicke parameter \cite{MC03}) 
in which the populations of both vibrational and internal levels oscillate.
A well known family corresponds to the ``motional sidebands'' 
which appear as avoided crossings in the dressed level 
structure with respect to laser-ion detuning for weak-fields. 
Less studied are the vibronic resonances for zero detuning 
and varying laser intensity \cite{MC99,JPK00,APS03}.
They may also be identified as avoided crossings 
with respect to the optically on-resonance Rabi frequency \cite{LRM06}.

It is in fact possible to identify and characterize  
all the above resonances with the same perturbative scheme \cite{APS03}, including 
moreover generalized resonances with respect to the detuning-adapted
Rabi frequency. 
Our first contribution in this paper has been to simplify the treatment originally
proposed by Anniello et al. \cite{APS03}. 
Our approach is based on a straightforward decomposition 
of the Hamiltonian (\ref{H}) into a zeroth order,
``semidressed'' term corresponding to 
$H$ with zero Lamb-Dicke parameter, and a coupling term responsible
for vibrational coupling.   
The resonances are at the degeneracies of a semidressed Hamiltonian
(in the $\{\Delta, \Omega_R\}$-plane on concentric circles where 
the detuning-adapted Rabi frequency is a multiple of the vibrational 
quantum),  
and the energy splittings at the avoided crossings of the dressed levels 
are estimated from degenerate perturbation theory. 
These splittings determine the resonance isolation, its width, 
and the oscillation frequency that gives the transition speed, 
all being crucial factors in applications for quantum gates or
state preparation.    

The formalism is very easy to adapt to hard-wall traps
so that the results for harmonic and square traps can be
compared. This comparison is our 
second main contribution. The main result is that, 
to leading order in $\eta$, the carrier transition, and vibronic 
resonances implying the interchange of 
three or more vibrational quanta are more effective (i.e., faster, 
more strongly coupled) for the hard wall trap. 
Explicit expressions have been provided and possible applications
of this intriguing behaviour are left for future research.

\begin{acknowledgments}
We thank J. I. Cirac and M. B. Plenio for useful comments 
on a preliminary version of this work presented in the 
XI edition of the workshop ``Time in Quantum Mechanics'' in La Laguna. 
This work has been supported by Ministerio de Educaci\'on y Ciencia
(BFM2003-01003),
and UPV-EHU (00039.310-15968/2004).\\
\end{acknowledgments}
%

%
%
%
%
%
\appendix
\section{Coupling Strengths in the harmonic potential}
\label{rabi_calculation}

The coupling strength, or effective Rabi frequency, is given by 
\begin{eqnarray}
\Omega_{nn'}&=&\Omega_{n'n}=\Omega_R\la n|e^{i\eta(a+a^\dag)}|n'\ra
\nonumber\\
&=&\Omega_R\int_{-\infty}^{\infty} dx \varphi_{n}(x)\varphi_{n'}(x)e^{ik_Lx},
\end{eqnarray}
where the $\varphi_n(x)$ are the normalized eigenfunctions of the harmonic potential. This integral gives \cite{WI79}
\begin{eqnarray}
\Omega_{nn'}\!=\!\Omega_{n'n}\!=\!
\Omega_R e^{-\eta^2/2}\left(i\eta\right)^{|n-n'|}\!\sqrt{\frac{n_{_<}!}{n_{_>}!}}
L_{n_{_<}}^{|n-n'|}(\eta^2)
\end{eqnarray}
where $n_{_{<}}$ ($n_{_>}$) is the lesser (greater) of $n$ and $n'$ and 
$L_n^{\alpha}$ are the generalized Laguerre functions, defined by \cite{Grandsteyn}
\begin{equation}
L_n^{\alpha}(X)=\sum_{k=0}^n (-1)^k \left( \begin{array}{c}
n+\alpha\\
n-k 
\end{array} \right)
\frac{X^k}{k!}.
\end{equation}
%
%
The leading term is given by \cite{wineland98}
\begin{eqnarray}
\Omega_{nn'}&=&\Omega_R(i\eta)^{|n-n'|}\frac{\sqrt{n_{_>}!/n_{_{<}}!}}{|n-n'|!}.
\end{eqnarray}
%
%
%

%
%
%
\section{Coupling Strengths in the square potential}
\label{square_potential}
For a hard wall trap of width $a$,
\begin{eqnarray}
V(x)=
\left\{ \begin{array}{cc}
0 & \textrm{if $|x|<\frac{a}{2}$}\nonumber\\
\hspace{1cm}\\
\infty & \textrm{if $|x|>\frac{a}{2}$}
                        \end{array}\right.,
\end{eqnarray}
the coupling strengths are given by the integral
\begin{equation}
\chi_{nn'}=\Omega_R\la \phi_n|e^{ik_L x}|\phi_{n'}\ra=\Omega_R\!\int_{-\frac{a}{2}}^{\frac{a}{2}} dx 
\phi_n(x) \phi_{n'}(x) e^{ ik_Lx},
\end{equation}
where $\phi_n(x)$ are the normalized eigenfunctions of the infinite well potential,
\begin{equation}
\phi_n(x)=\la x|\phi_n\ra=
\left\{ \begin{array}{cc}
\sqrt{\frac{2}{a}}\cos{\frac{n\pi}{a}x} & \textrm{if $n$ odd}\nonumber\\
\hspace{1cm}\\
\sqrt{\frac{2}{a}}\sin{\frac{n\pi}{a}x} & \textrm{if $n$ even}
\end{array}\right. ,
\end{equation}
satisfying the boundary conditions $\phi_n(-\frac{a}{2})=\phi_n(\frac{a}{2})=0$. This integral gives
\begin{equation}
\chi_{nn'}=-(k_La) \Omega_R i^{|n-n'|}\frac{8 nn'\pi^2}{D}
\left\{ \begin{array}{cc}
\sin{\frac{k_La}{2}} &\textrm{if $n-n'$ even}\nonumber\\
\hspace{1cm}\\
\cos{\frac{k_La}{2}}& \textrm{if $n-n'$ odd}\nonumber\\
                        \end{array}\right.,
\end{equation}
where $D$ is a common denominator given by
\begin{eqnarray}
D&=&\left[(k_La)^2-(n'-n)^2\pi^2\right]\left[(k_La)^2-(n'+n)^2\pi^2\right]\nonumber
\end{eqnarray}
Note that, as in the harmonic case, 
these coupling strenghts are real when connecting motional levels of the same parity,  and purely imaginary otherwise.

We define the LD parameter for this type of potential in analogy with the harmonic case, in which 
the LD parameter could be written as $\eta=\frac{\hbar k_L}{p_h}$ with $p_h$ being the quasimomentum 
in $\hbar\omega_T=\frac{p_h^2}{2m}$. For a hard-wall potential
$\eta$ can be written as $\eta=\frac{\hbar k_L}{p_{hw}}$ with $p_{hw}$ being the quasi-momentum unit in the energy eigenstates, i.e., $p_{hw}=\hbar\pi/a$, so that the LD parameter is defined as $\eta:=k_La/\pi$. In terms
of this LD parameter the coupling strength reads
\begin{equation}
\chi_{nn'}=-\eta \Omega_R i^{|n-n'|}\frac{8 nn'}{\pi\tilde D}
\left\{ \begin{array}{cc}
\sin{\frac{\eta\pi}{2}} &\textrm{if $n-n'$ even}\nonumber\\
\hspace{1cm}\\
\cos{\frac{\eta\pi}{2}}& \textrm{if $n-n'$ odd}\nonumber\\
                        \end{array}\right.
\end{equation}
with
\begin{equation}
\tilde D=\left[\eta^2-(n'-n)^2\right]\left[\eta^2-(n'+n)^2\right].\nonumber
\end{equation}

In the LD regime, the coupling strength to leading order in $\eta$ is given by
%
\begin{equation}
\chi_{nn'}\approx-\Omega_R \frac{4nn'i^{|n-n'|}}{(n-n')^2(n+n')^2}
\left\{ \begin{array}{cc}
\eta^2&\textrm{if $n-n'$ even}\nonumber\\
\hspace{1cm}\\
2\eta/\pi & \textrm{if $n-n'$ odd}\nonumber\\
                        \end{array}\right..
\end{equation}
\vspace{1.5cm}


\end{document}